\documentclass[twocolumn,showpacs,showkeys]{revtex4}

\usepackage{amsfonts}
\usepackage{amsmath}
\usepackage{graphicx}

\begin{document}


\title{A proposal of foundation of spacetime geometry}


\author{Romualdo Tresguerres}
\email[]{romualdotresguerres@yahoo.es}
\affiliation{Instituto de F\'isica Fundamental\\
Consejo Superior de Investigaciones Cient\'ificas\\ Serrano 113 bis, 28006 Madrid, SPAIN}

\date{\today}

\begin{abstract}
A common approach to metric-affine, local Poincar\'e, special-relativistic and Galilei spacetime geometry is developed. Starting from an affine composite bundle, we introduce local reference frames and their evolution along worldlines and we study both, absolute and relative simultaneity postulates, giving rise to alternative concepts of spacetime. In particular, the construction of the Minkowski metric, and its required invariance, allows either to reorganize the original affine bundle as a metric-affine geometry with explicit Lorentz symmetry, or to restrict it to a Poincar\'e geometry, both of them constituting the background of a wide class of gauge theories of gravity.
\end{abstract}

\maketitle

\section{Introduction}

Spacetime geometry is the primary framework underlying any physical theory. Newtonian mechanics rests on the definitions of absolute space and time, Einstein's Special Relativity provides a basis for the consistent treatment of electromagnetism and dynamics, as much as for the description of fundamental forces -up to gravity- in the Standard Model, and General Relativity (GR), as the theory of gravitational phenomena, interprets them in geometrical terms. Quantum Mechanics also evolved from non-relativistic to relativistic formulations, and even Thermodynamics presupposes a concept of space and time \cite{Tresguerres:2013rnr}.

The lack of a satisfactory quantization procedure for GR, as much as the aim to get a unified picture of all interactions, justifies the search for an alternative gauge theory of gravity, based on the local realization of a spacetime group. Poincar\'e gauge theories and Metric-Affine Gravity \cite{Hehl:1974cn}-\cite{Ali:2007hu}, containing GR as a particular case, are examples of such a gauge-theoretical formulation of gravitational forces, similar to that of the Standard Model. The purpose of the present work is to study the geometric foundations of spacetime gauge theories of this kind or, expressing the same thing in a more far-reaching language, to investigate the general foundations of the geometry of physical spacetime.

Roughly speaking, geometries are structured manifolds. Felix Klein in his Erlangen Program \cite{Sharpe:1996} characterized them as manifolds endowed with a group action, being geometric objects identified with the invariants preserved under group transformations. Moreover, the Riemannian geometry of standard general-relativistic spacetime consists of a metric manifold $(M\,,g)$, where Levi-Civita connections and the corresponding curvature are constructed from the metric. In what follows, we assume the general treatment of geometry introduced by \'Elie Cartan \cite{Cartan}. According to \cite{Sharpe:1996}, Cartan geometries are modeled on Klein's (flat) geometries, generalizing them with the help of connections in the same way as (curved) Riemannian geometry generalizes Euclidean geometry. A Cartan geometry is suitably formalized, in analogy to physical gauge theories, as a principal fiber bundle of a given Lie group $G$ (with a closed subgroup $H$) where a certain connection is defined. For details, see \cite{Sharpe:1996}, pg.184. In view of previous results \cite{Tresguerres:2002uh} \cite{Tresguerres:2012nu}, in the present paper we propose a reformulation of the fundamental structure of Cartan geometries as that of composite fiber bundles, identified by us as the suitable geometric framework to deal with nonlinear realizations of gauge theories \cite{Coleman:1969sm}-\cite{Tresguerres:2008jf}. The latter ones are relevant primarily in the context of spontaneous symmetry breaking mechanisms \cite{Tresguerres:2008jf}, but also for solving some difficulties concerning the gauge treatment of translational symmetry, for instance the seemingly unavoidable non-locality of such transformations, or the problematic geometric interpretation of translational connections, due to their inhomogeneous gauge transformations. (See Refs.\cite{Tresguerres:2002uh} \cite{Tresguerres:2012nu}.) The composite bundle approach provides a satisfactory framework, clarifying the gauge treatment of translations and allowing one to identify (nonlinear) translational connections as tetrads, that is, as the 1-form basis of the cotangent space of spacetime.

So, our starting point is an axiomatic presentation of the main elements of composite bundle affine geometry, summarizing results found in \cite{Tresguerres:2012nu} in the context of Poincar\'e gauge theories. We choose the parallelism preserving affine structure as fundamental, following Weyl \cite{Weyl} and Hehl \cite{Hehl:1995ue}, in view of its generality and simplicity, despite the possibility of deriving it from other hypotheses. Actually, Ehlers et al. \cite{Ehlers:1972}- \cite{Capozziello:2012eu} arrive at affine geometry from the requirement of compatibility of the more basic projective and conformal structures. The reason for them to proceed in this way is that they attempt to construct the geometry of spacetime deriving it from observational quantities such as light rays and freely falling particles. Instead, we postulate the affine background as an auxiliary formal framework previous to experience, whatever its ontological {\it status} may be, and we take from it the geometrical objects which are relevant for the operational description of positions, trajectories, relative motions, etc., constituting, say, the {\it observable} aspects of spacetime geometry emerging from the underlying theoretical domain.

In a next step, we consider the transition from affine to metric geometries, enlarging the original structure by introducing a metric whose form is determined with the help of suitable synchronization hypotheses. We follow two different constructive approaches, based on absolute and relative simultaneity postulates, leading respectively to Galilean geometry and to a metric spacetime endowed with a Minkowski metric. In metric spaces, the congruence of line elements and vectors defined at different points is guaranteed provided they can be transformed into each other by means of a suitable isometry group, which for the Minkowski metric is found to be the Poincar\'e group.

The paper is organized as follows. In Section II, the composite fiber bundle structure of affine geometry is introduced, together with its tangent and cotangent spaces, providing the background to define affine frames and coframes. Events are described not by coordinates of the base space, but by the components, referred to a local origin, of the geometric object which we call {\it position}. One of its components is identified as locally measurable {\it clock time}. Section III is devoted to worldlines. Related to them, a second kind of time, namely {\it parametric time}, responsible for evolution, is considered \cite{Rovelli:1990jm}-\cite{Rovelli:1995}. Parametric time evolution is studied in single worldlines as much as in mutually oblique ones (moving relatively to each other), making apparent the need for a synchronization criterion. In Section IV we deal with absolute simultaneity, characteristic for absolute space and time as much as for the relativistic Galilean spacetime of Newtonian mechanics. In Section V, Einstein's relative simultaneity is studied, leading to the construction of the Minkowski metric, invariant under local transformations of the Poincar\'e group, and we derive the corresponding (local) Lorentz transformations. We end with some comments on dynamics in Section VI and with the Conclusions, where we point out the consequences of our approach for the different concepts of spacetime derivable from it.

\section{Affine geometry in a composite bundle}

Our approach to Cartan geometries is based on composite fiber bundles, constructed according to Sardanashvily \cite{Sardanashvily}--\cite{Sardanashvily:2005mf} with the help of propositions 5.5 and 5.6 of Ref. \cite{Kobayashi:1963} as follows. Provided $\pi _{_{PM}}:P\rightarrow M$ is a principal fiber bundle whose structure group $G$ is reducible to a closed subgroup $H\subset G$, it is possible to rearrange the bundle as a composite manifold
\begin{equation}
\pi _{_{\Sigma M}}\circ\pi _{_{P\Sigma}}:P\rightarrow\Sigma\rightarrow M\,,\label{compbundle01}
\end{equation}
being $\Sigma =P/H$ an {\it intermediate space} such that
\begin{equation}
\pi _{_{P\Sigma}}:P\rightarrow\Sigma\label{compbundle02}
\end{equation}
is a principal subbundle of $P$ with structure group $H$, and
\begin{equation}
\pi _{_{\Sigma M}}:\Sigma\rightarrow M\label{compbundle03}
\end{equation}
is a $P$-associated bundle with typical fiber $G/H$ and structure group $G$. The projections in (\ref{compbundle01})-(\ref{compbundle03}) are related as $\pi _{_{\Sigma M}}\circ\pi _{_{P\Sigma}} =\pi _{_{PM}}$, and global sections $s_{_{M\Sigma}}: M\rightarrow\Sigma $ of (\ref{compbundle03}) exist, playing the role of Goldstone-like fields.

In particular, we construct an affine composite bundle geometry based on the affine group $A(4\,,\mathbb{R}) = GL(4\,,\mathbb{R}){\;{\rlap{$\subset$}\times}\;} T^4$ consisting in the semi-direct product of the general linear group $GL(4\,,\mathbb{R})$ and the commutative group $T^4 = \mathbb{R}^4$ of spacetime translations in four dimensions. In the scheme (\ref{compbundle01})-(\ref{compbundle03}), we take $G$ to be the affine group $A(4\,,\mathbb{R})$, with $H$ as its general linear subgroup $GL(4\,,\mathbb{R})$, assuming the 4-dimensionality of the base space $M$, to which all geometric quantities are to be pulled back.

Given the composite bundle (\ref{compbundle01}) as a manifold, one can study the tangent and cotangent bundles of its different pieces separately. In \cite{Tresguerres:2012nu} we paid special attention to the tangent and cotangent bundles $T(\Sigma )$ and $T^*(\Sigma )$ of (\ref{compbundle03}) in order to introduce spacetime frames and coframes. Let us present a summary of the main results, expressed in a simplified notation where we do not distinguish between the quantities defined on $\Sigma$ and the corresponding ones pulled back to $M$. The interested reader is referred to \cite{Tresguerres:2012nu} for a detailed deduction (performed there for Poincar\'e geometry, but immediately generalizable to affine geometry).

\subsection{Local reference frames and positions}

In the vertical sector of the tangent space $T(\Sigma )$, one constructs a vector basis $\{ e_\alpha\}$ ($\alpha =0\,,1\,,2\,,3$) and, by adding a point $\mathfrak{o}\in T(\Sigma )$, one completes \cite{Kobayashi:1963} the affine frames
\begin{equation}
e_A = \binom{e_\alpha }{l^{-1}\mathfrak{o}\,}\,,\label{5dpoinc72a}
\end{equation}
($A=0\,,1\,,2\,,3\,,5$), which we identify \cite{Tresguerres:2012nu} as the constituents of an affine frame bundle transforming under the 5x5 matrix representation of the affine group (see Appendix A), with $l^{-1}\mathfrak{o}$ standing for the fifth affine vector component. The tangent bundle $T(\Sigma )$ can then be regarded as an associated vector bundle of the principal bundle of local frames (\ref{5dpoinc72a}) with the affine group as its structure group.

In the composite fiber bundle approach considered by us (constituting a natural framework for nonlinear realizations) \cite{Tresguerres:2002uh} \cite{Tresguerres:2012nu}, frames (\ref{5dpoinc72a}) are replaced by the modified ones
\begin{equation}
\widehat{e}_A = (b^{-1})_A{}^B\,e_B = \binom{e_\alpha }{l^{-1}\,p\,}\,,\label{5dpoinc72b}
\end{equation}
built from (\ref{5dpoinc72a}) with (\ref{5dtransmat}). Affine gauge transformations to be considered in the next paragraph make apparent that the redefined frames (\ref{5dpoinc72b}), with invariant fifth component
\begin{equation}
p = \mathfrak{o} +\xi ^\alpha\,e_\alpha\,,\label{5dpoinc73}
\end{equation}
behave as $GL(4\,,\mathbb{R})$ representation fields, as is characteristic for the nonlinear realization of the affine group with $H=GL(4\,,\mathbb{R})$.

The quantity (\ref{5dpoinc73}) replacing in (\ref{5dpoinc72b}) the origin present in (\ref{5dpoinc72a}), consists of the origin $\mathfrak{o}$ plus vector contributions whose components are the translational Goldstone fields $\xi ^\alpha$ introduced in (\ref{5dtransmat}) as coset parameters of $G/H$. These fields $\xi ^\alpha$ play the role of a certain kind of {\it coordinates}, determining the relative position of events with respect to the origin $\mathfrak{o}$ and the basis vectors $\{e_\alpha\}$ of the frame (\ref{5dpoinc72a}). We call a world-point described by (\ref{5dpoinc73}) a {\it position}, and we assume the coordinate-like position components $\xi ^\alpha$ of events to be locally measurable by means of clocks and rods. In particular, $\xi ^0$ represents {\it clock time} as read out from a {\it local clock}, the latter consisting for instance in an oscillating device whose regular frequency is guaranteed by accepted physical laws, say those of Quantum Mechanics. The {\it position} points (\ref{5dpoinc73}) play a fundamental role in what follows.

\subsection{Gauge transformations, connections and coframes}

The frames (\ref{5dpoinc72a}) or (\ref{5dpoinc72b}) defined in the tangent space $T(\Sigma )$ of the {\it intermediate space} $\Sigma$ of the composite bundle can experience two kinds of geometric motions. On the one hand, the gauge transformations of the different pieces of (\ref{5dpoinc73}) under the affine group are found to be
\begin{eqnarray}
\delta e_\alpha &=& \beta _\alpha{}^\beta\,e_\beta\,,\label{5dpoinc74}\\
\delta\mathfrak{o} &=& \epsilon ^\alpha\,e_\alpha \,,\label{5dpoinc75}\\
\delta\xi ^\alpha &=&-\xi ^\beta\beta _\beta{}^\alpha -\epsilon ^\alpha\,,\label{5dpoinc76}\\
\delta p &=& 0\,.\label{5dpoinc77}
\end{eqnarray}
They are vertical along fibres $G/H$, involving the gauge group parameters $\epsilon ^\alpha$ and $\beta _\alpha{}^\beta$, translating, rotating deforming or boosting frames with respect to each other. According to (\ref{5dpoinc74})--(\ref{5dpoinc77}), the basis vectors $e_\alpha$ transform as $GL(4\,,\mathbb{R})$ vectors, while translations and general linear transformations of both, the origin and the translational coordinate-like fields, occur in such a way that the position (\ref{5dpoinc73}) is left invariant in any reference frame. Thus, as previously announced, the redefined frame (\ref{5dpoinc72b}) transforms under the whole affine group as a $GL(4\,,\mathbb{R})$ object. On the other hand, the lateral (or horizontal) displacements of the same quantities read
\begin{eqnarray}
\nabla e_\alpha &=& \Gamma _\alpha{}^\beta\,e_\beta\,,\label{5dpoinc78}\\
\nabla \mathfrak{o} &=& {\buildrel (T)\over{\Gamma ^\alpha}}\,e_\alpha \,,\label{5dpoinc79}\\
\nabla p &=& \nabla (\mathfrak{o} +\xi ^\alpha\,e_\alpha )\nonumber\\
&=& \vartheta ^\alpha\,e_\alpha \,.\label{5dpoinc80}
\end{eqnarray}
They are moved on $\Sigma$ by the operator $\nabla$ (the so called {\it affine connection}), whose action on the frames (\ref{5dpoinc72a}) or (\ref{5dpoinc72b}) makes connections emerge, being $\Gamma _\alpha{}^\beta $ the $GL(4\,,\mathbb{R})$ and ${\buildrel (T)\over{\Gamma ^\alpha}}$ the (linear) translational connections respectively. (Connections define horizontality in a bundle. Those appearing in (\ref{5dpoinc78})-(\ref{5dpoinc80}) are derived in \cite{Tresguerres:2012nu} from the ones of the affine principal bundle.) While vertical (gauge) transformations merely modify the point of view, lateral ones induce effective changes (\ref{5dpoinc78}) and (\ref{5dpoinc80}) of the basis vectors and of the position $p$ defined in (\ref{5dpoinc73}) respectively. The modified (nonlinear) translational connections introduced in (\ref{5dpoinc80}) have the structure
\begin{equation}
\vartheta ^\alpha := D\xi ^\alpha +{\buildrel (T)\over{\Gamma ^\alpha}}\,,\label{partlight01}
\end{equation}
with
\begin{equation}
D\xi ^\alpha := d\xi ^\alpha + \Gamma _\beta{}^\alpha\,\xi ^\beta \,.\label{partlight02}
\end{equation}
Being the gauge transformations of the connections, as found in \cite{Tresguerres:2012nu},
\begin{equation}
\delta\Gamma _\alpha{}^\beta =D\beta _\alpha{}^\beta\label{varlorconn}
\end{equation}
and
\begin{equation}
\delta {\buildrel (T)\over{\Gamma ^\alpha}}=-{\buildrel (T)\over{\Gamma ^\beta}}\beta _\beta{}^\alpha +D\epsilon ^\alpha\,,\label{vartransconn}
\end{equation}
it follows that the variation of (\ref{partlight01}) reads
\begin{equation}
\delta\vartheta ^\alpha =-\vartheta ^\beta\,\beta _\beta{}^\alpha\,.\label{vartetrad}
\end{equation}
Thus, the nonlinear translational connections (\ref{partlight01}) transform as covectors. Moreover, as proved in \cite{Tresguerres:2012nu}, they satisfy
\begin{equation}
e_\alpha\rfloor \vartheta ^\beta = \delta _\alpha ^\beta\,,\label{dualitycond}
\end{equation}
so that $\{\vartheta ^\alpha\}$ can play the role of a set of tetrads, that is, of a suitable 1-form basis (coframe) of the cotangent space $T^*(\Sigma )$, dual to the vector basis $\{ e_\alpha\}$ of $T(\Sigma )$.

In order to complete the number of relevant geometrical quantities, curvature and torsion are defined in Appendix B.

\section{Worldlines}

We will describe particle trajectories and light rays as worldlines (curves) on the intermediate space $\Sigma$ rather than on the base space $M$. Therefore, in the present section first we briefly collect the main definitions and results concerning curves in general, following \cite{Nakahara}, and next we study the interplay between worldlines and the quantities introduced in previous section, mainly positions, frames and coframes, characteristic for the geometric structure of $\Sigma$.

\subsection{Curves, tangent vectors and Lie derivatives}

A curve in an $n$-dimensional manifold $M$ is a map $\gamma : I\rightarrow M$ from an interval $I\subset\mathbb{R}$ to $M$, the map image $\gamma (\tau )\subset M$ being a smooth one-dimensional submanifold parametrized by the values of a real variable $\tau$. By means of a suitable coordinate map $\varphi : M\rightarrow\mathbb{R}^n$, we get a coordinate representation of the curve $\gamma (\tau )$ as $\varphi\left(\gamma (\tau )\right) =\left\{ x^i(\tau )\right\}\in\mathbb{R}^n$.

The vector $u$ tangent to a curve at a given point $\gamma (0)$ is defined with the help of an arbitrary function $f: \gamma (\tau )\rightarrow\mathbb{R}$. In terms of the curve coordinates, the function becomes $f\left( \gamma (\tau )\right) = f\circ\varphi ^{-1}\left( x^i(\tau )\right)$, so that one can evaluate its rate of change along the curve at $\tau =0$ to be
\begin{eqnarray}
\left.\dfrac{d f\left(\gamma (\tau )\right)}{d\tau}\right|_{\tau =0} &=& \left.\dfrac{d x^i(\tau )}{d\tau}\,\partial _i\left( f\circ\varphi ^{-1}\right)\right|_{\tau =0}\nonumber\\
&=:& \left.u^i\left( x^j(\tau )\right)\partial _i\left( f\circ\varphi ^{-1}\right)\right|_{\tau =0}\nonumber\\
&=& u\rfloor d\left( f\circ\varphi ^{-1}\right)\,.\label{curv01}
\end{eqnarray}
The r.h.s. of (\ref{curv01}) is a directional derivative of the function $f\left(\gamma (\tau )\right)$ along $\gamma (\tau )$, with the tangent vector $u$ pointing in the direction of the curve at $\gamma (0)$.

Conversely, given a vector field $u$, the condition
\begin{equation}
\dfrac{d x^i(\tau )}{d\tau} = u^i\left( x^j(\tau )\right)\label{curv02}
\end{equation}
read out from (\ref{curv01}) determines a family of integral curves $\gamma (\tau )$ with coordinates $\left\{ x^i(\tau )\right\}$ having $u$ as their tangent vector field. According to the theorem of existence and uniqueness of solutions to first order ordinary differential equations with given initial conditions, a unique solution of Eq.(\ref{curv02}) exists for each initial value, so that the curve $\gamma _p(\tau )$ passing through an arbitrary point $p=\gamma _p(0)\in M$ is unique. The totality of the non-intersecting space-filling integral curves is called the congruence of curves generated by $u$. A congruence can be represented as a flow defined as follows. Each representative $\gamma _p(\tau )$ of the congruence of curves is a map $\gamma _p : I\rightarrow M$ depicting a curve through $p\in M$ as $\tau\rightarrow \gamma _p(\tau )$. The map $\Phi : I\times M\rightarrow M$ such that $(\tau\,,p)\rightarrow\Phi (\tau\,,p) =\gamma _p(\tau )$, which takes into account all possible points $p\in M$, is called a flow generated by $u$.

The coordinates of $\gamma _p(\tau +\sigma ) =\Phi (\tau +\sigma\,,p)$ and those of $\gamma _{\tiny\mbox{$\gamma _p(\sigma )$}}(\tau ) =\Phi (\tau\,,\Phi (\sigma\,,p))$, both satisfy Eq.(\ref{curv02}) with equal initial condition (see \cite{Nakahara}, pg.151), so that, in view of the uniqueness theorem of ordinary differential equations, both curves are the same, that is
\begin{equation}
\gamma _{\tiny\mbox{$\gamma _p(\sigma )$}}(\tau ) = \gamma _p(\tau +\sigma )\,,\label{curv04}
\end{equation}
with initial conditions
\begin{equation}
\gamma _{\tiny\mbox{$\gamma _p(\sigma )$}}(0) = \gamma _p(\sigma )\,.\label{curv06}
\end{equation}
The integral curves $\gamma _p(\tau )$ found by solving (\ref{curv02}) allow one to define also
\begin{equation}
\phi _\tau (p):=\gamma _p(\tau )\,,\label{curv07}
\end{equation}
that is, diffeomorphisms $\phi _\tau : M\rightarrow M$ on the manifold $M$ displacing $p\in M$ along the curves. In terms of (\ref{curv07}), one reformulates (\ref{curv04})-(\ref{curv06}) as
\begin{eqnarray}
\phi _\tau \circ \phi _\sigma &=& \phi _{\tau +\sigma}\,,\label{curv07bis}\\
\phi _0 \circ \phi _\sigma &=& \phi _\sigma\,.\label{curv08}
\end{eqnarray}
From (\ref{curv07bis}) follow the properties
\begin{eqnarray}
\phi _\tau \circ \phi _{-\tau} &=& \phi _0\,,\label{curv09}\\
(\phi _\tau \circ \phi _\sigma )\circ \phi _\nu &=& \phi _\tau \circ (\phi _\sigma\circ \phi _\nu )\,,\label{curv10}
\end{eqnarray}
proving that the diffeomorphisms $\phi _\tau$ constitute a commutative one-parameter group of transformations (along the curve), since (\ref{curv07bis}) guarantees closure and commutativity of the group operation, (\ref{curv08}) and (\ref{curv09}) show the existence of an identity element and of inverse elements respectively, and (\ref{curv10}) expresses associativity. In view of (\ref{curv07}), the coordinates of the transformed point $\phi _\tau (p)$ coincide with those of the point $\gamma _p(\tau )$ of the curve, that is
\begin{equation}
\varphi\left(\phi _\tau (p))\right) = \varphi\left(\gamma _p(\tau )\right) = \left\{ x_p^i(\tau )\right\}\,,\label{curv10bis}
\end{equation}
and for infinitesimal transformations, using (\ref{curv02}), they are found to expand as
\begin{equation}
x_p^i(\epsilon )\approx x_p^i(0) +\epsilon\, u^i( x_p^j(0))\,,\label{curv11}
\end{equation}
showing that $\phi _\tau$ transformations are generated by the tangent vector field $u$.

With the help of the diffeomorphisms $\phi _\tau$, one can construct Lie derivatives, enabling to compare vectors, differential forms and general tensor fields defined at different points of a curve. Indeed, given a tensor field with values $\mathbb{T}(p)$ and $\mathbb{T}(\phi _\tau (p))$ at neighboring points, use is made of an induced map $\phi _\tau ^*$ to drag $\mathbb{T}(\phi _\tau (p))$ back to the point $p$, where its comparison with $\mathbb{T}(p)$ defines the Lie derivative of the tensor with respect to the vector field $u$ tangent to the curve as
\begin{equation}
{\it{l}}_u \mathbb{T}:= \lim_{\tau\to 0}\,{1\over{\tau}}\left[\phi _\tau ^*\,\mathbb{T}(\phi _\tau (p)) -\mathbb{T}(p)\,\right]\,,\label{curv12}
\end{equation}
(see \cite{Hehl-and-Obukhov}). The Lie derivative measures the change of $\mathbb{T}$ induced by $\phi _\tau$ along the curve $\gamma _p(\tau )$. In particular, the Lie derivative of a vector $X$ can be proven to be
\begin{equation}
{\it{l}}_u X = \left[ u\,,X\right]\,,\label{curv13}
\end{equation}
and that of a $p$-form
\begin{equation}
{\it{l}}_u\alpha = u\rfloor d\alpha + d\,(u\rfloor\alpha\,)\,.\label{Liederdef}
\end{equation}
In summary, we have introduced curves $\gamma (\tau )$, tangent vector fields $u$ of the curves and Lie derivatives measuring the change of geometric objects along the curves. Given a nowhere vanishing vector field $u$, its integral curves constitute a congruence of non-intersecting trajectories passing through neighboring points. Reciprocally, the congruence determines $u$ as its tangent vector field. The congruence of curves also defines a flow, consisting in a transformation of the manifold into itself (generated by the vector field $u$) along the curves. The congruence of curves, its tangent vector field and the flow generated by the latter reciprocally imply each other \cite{Crampin:1986}.

\subsection{Parametric time and spacetime foliation}

Having Eq.(\ref{curv01}) in view, let us consider as a particular case the function assigning to any point $\gamma (\tau)$ of the curve the value of its parameter at that point, that is
\begin{equation}
f\left(\gamma (\tau )\right) = f\circ\varphi ^{-1}(x^i(\tau )) = \tau (x^i(\tau ))\,.\label{spfol01}
\end{equation}
For this choice, one gets trivially
\begin{equation}
\dfrac{d f\left(\gamma (\tau )\right)}{d\tau} =\dfrac{d\tau}{d\tau} =1\,,\label{spfol02}
\end{equation}
so that from (\ref{curv01}) with (\ref{spfol01}) and (\ref{spfol02}) follows
\begin{equation}
u\rfloor d\tau =1\,,\label{spfol03}
\end{equation}
allowing the vector field $u$ to be expressed as
\begin{equation}
u=\partial _\tau\,.\label{spfol04}
\end{equation}
Notice that, in view of (\ref{Liederdef}), the condition (\ref{spfol03}) defining $u$ in terms of $\tau$ is equivalent to
\begin{equation}
{\it l}_u\,\tau =1\,.\label{partlight08}
\end{equation}
In the following we identify the parameter $\tau$ as a certain time variable, and we assign to it dimensions of time. The time vector field $u$ related to it through (\ref{spfol03}), being the tangent vector of a congruence of curves, allows one to formalize time evolution of any physical quantity represented by a $p$-form $\alpha$ as its Lie derivative (\ref{Liederdef}) along worldlines.

According to the form in which {\it parametric time} $\tau$ is introduced, associated to particular curves (or to congruences of curves), it clearly differs from Newtonian universal time. Moreover, in principle it neither has to do with clock time understood as the locally measurable $\xi ^0$ component of (\ref{5dpoinc73}), since no operational way to relate both time variables is defined for the moment. However, later we will introduce two alternative assumptions for the measurement of parametric time in terms of clock time, corresponding to absolute and relative conceptions of simultaneity respectively.

The 1-form $\omega = d\tau $ in (\ref{spfol03}) trivially satisfies the Frobenius' foliation condition $\omega\wedge d\omega =0$. Thus, relatively to the direction of the time vector $u$, any $p$-form $\alpha$ can be decomposed into longitudinal and transversal contributions \cite{Hehl-and-Obukhov} as
\begin{equation}
\alpha = d\tau\wedge\alpha _{\bot} +\underline{\alpha}\,,\label{foliat1}
\end{equation}
being the longitudinal piece
\begin{equation}
\alpha _{\bot} := u\rfloor\alpha\label{long-part}
\end{equation}
the projection of $\alpha$ along $u$, while the transversal component
\begin{equation}
\underline{\alpha}:= u\rfloor ( d\tau\wedge\alpha\,)\label{trans-part}
\end{equation}
is orthogonal to the former as a spatial projection. The foliation of the exterior derivative of a form (\ref{foliat1}) reads
\begin{equation}
d\,\alpha = d\tau\wedge\bigl(\,{\it{l}}_u\underline{\alpha} -\,\underline{d}\,\alpha _{\bot}\,\bigr) +\underline{d}\,\underline{\alpha }\,,\label{derivfoliat}
\end{equation}
where the longitudinal part is expressed in terms of the Lie derivative (\ref{Liederdef}) of (\ref{trans-part}) and of the spatial differential $\underline{d}$ of (\ref{long-part}).

\subsection{Evolution of positions}

Let us consider worldlines defined in the intermediate space $\Sigma$. There we introduce parametric time evolution as induced by the affine evolution operator $\nabla _u$ representing the action of $\nabla$ (see (\ref{5dpoinc78})-(\ref{5dpoinc80})) along curves $\gamma (\tau )$ with tangent time vector $u$. Taking into account Eq.(\ref{5dpoinc80}) of the displacement of the position $p$ by the operator $\nabla\,$, we define the related action of $\nabla _u$ on $p$ as
\begin{equation}
\nabla _u p := \bigl(\,u\rfloor\vartheta ^\alpha\bigr) e_\alpha =: u^\alpha e_\alpha\,,\label{vel01}
\end{equation}
showing the effect of carrying $p$ along the flow generated by $u$. Recalling the tetrad structure (\ref{partlight01}) with (\ref{partlight02}) and definition (\ref{Liederdef}), we find
\begin{equation}
u^\alpha := u\rfloor\vartheta ^\alpha = {\cal \L\/}_u\xi ^\alpha +{\buildrel (T)\over{\Gamma _{\bot}^\alpha}}\,,\label{partlight03}
\end{equation}
where we introduce the covariant generalization of definition (\ref{Liederdef}) of Lie derivatives as
\begin{equation}
{\cal \L\/}_u\xi ^\alpha := u\rfloor D\xi ^\alpha = {\it{l}}_u\xi ^\alpha + {\Gamma _{\bot}}_\beta{}^\alpha\,\xi ^\beta\,,\label{partlight04}
\end{equation}
involving covariant differentials instead of ordinary ones \cite{Hehl:1995ue}. The vector components (\ref{partlight03}) of $u= u^\alpha e_\alpha$ as found in (\ref{vel01}) (describing the same vector $u$ as (\ref{spfol04})) are interpreted as four-velocity components. They consist, apart from connections, of the Lie derivatives ${\it{l}}_u\xi ^\alpha$ measuring the evolution of the coordinate-like position fields with respect to parametric time. An observer located in the vicinity of the origin of (\ref{5dpoinc73}) experiences the displacement of position $p$ along a worldline as a succession of changing values of the position components $\xi ^\alpha$ relative to the local reference frame. (That is what {\it observable motion} is about.)

Acting twice with $\nabla _u $ on $p$, we get the covariant acceleration which, in view of (\ref{vel01}), reads
\begin{eqnarray}
\nabla _u \nabla _u p &=& \nabla _u u\nonumber\\
&=& {\cal \L\/}_u u^\alpha \,e_\alpha\,,\label{eqsmot}
\end{eqnarray}
where the acceleration components can be expressed, alternatively, as
\begin{eqnarray}
{\cal \L\/}_u u^\alpha &:=& {\it{l}}_u u^\alpha +\Gamma  _{\bot\beta}{}^\alpha\,u^\beta \nonumber\\
&=& {\cal \L\/}_u\,{\cal \L\/}_u\,\xi ^\alpha + {\cal \L\/}_u\,{\buildrel (T)\over{\Gamma _{\bot}^\alpha}}\,.\label{acceleration}
\end{eqnarray}
In Eq.(\ref{eqsmot}), $\nabla _u \nabla _u p$ measures the rate of change of the vector $\nabla _u p$ (equal to the tangent vector $u$) in the direction of $u$. Vanishing of (\ref{eqsmot}) and thus of (\ref{acceleration}) gives rise to autoparallel (inertial) motion.

\subsection{Evolution along mutually oblique worldlines (relative motion)}

Given a congruence of curves having $u$ as tangent vector, let us consider the displacement of (\ref{5dpoinc73}) along a curve with tangent vector $w$ oblique to the congruence. In analogy to (\ref{vel01}), we find
\begin{equation}
\nabla _w p = w^\alpha e_\alpha\,,\label{wvector01}
\end{equation}
where
\begin{equation}
w^\alpha := w\rfloor\vartheta ^\alpha\,.\label{wvector02}
\end{equation}
In order to compare the mutually oblique fourvelocities (\ref{wvector02}) and (\ref{partlight03}) to each other, let us apply the decomposition (\ref{foliat1}) to the tetrads (\ref{partlight01}) as
\begin{equation}
\vartheta ^\alpha = d\tau\,u^\alpha + \underline{\vartheta}^\alpha\,,\label{partlight09}
\end{equation}
(where $u^\alpha$ is given by (\ref{partlight03})), and replace (\ref{partlight09}) in (\ref{wvector02}). So we find
\begin{equation}
w^\alpha = (w\rfloor d\tau )\,u^\alpha +  w\rfloor\underline{\vartheta}^\alpha\,,\label{partlight05}
\end{equation}
in terms of the projections of $w$ on the longitudinal and transversal parts of (\ref{partlight09}) relative to $u$. By defining the transversal velocity contribution as
\begin{equation}
v^\alpha := \,{{w\rfloor\underline{\vartheta}^\alpha}\over{w\rfloor d\tau }}\,,\label{partlight06}
\end{equation}
we rewrite (\ref{partlight05}) in the form
\begin{equation}
w^\alpha = (w\rfloor d\tau )\,\left(\,u^\alpha + v^\alpha\,\right)\,.\label{partlight07}
\end{equation}
Contrary to (\ref{spfol03}), evaluated along a single worldline, the quantity $(w\rfloor d\tau )$ involves parametric time $\tau$ of different worldlines crossed by $w$, the latter ones belonging to the congruence with tangent vector $u$, so that $d\tau$ is to be understood as the infinitesimal limit of the difference $\tau _B -\tau _A$ between time parameter values of separate curves A and B. Thus, relative speeds along mutually oblique paths are in principle not fully determined, mainly due to the lack of a synchronization criterion allowing one to evaluate $(w\rfloor d\tau )$ in (\ref{partlight05}) (or in (\ref{partlight07}) with (\ref{partlight06})). In what follows, we will consider two possible solutions to this difficulty. One of them consists in the acceptance of absolute simultaneity as in Newtonian mechanics, and the other one rests on Einstein's definition of relative simultaneity.

In our approach, synchronization of distant events, as required by the local character of time recognized by Einstein, involves the two kinds of local time previously introduced, namely {\it clock time} $\xi ^0$ (the time component of (\ref{5dpoinc73}) measurable by local clocks) and parametric time $\tau$ responsible for evolution (even of $\xi ^0$) along worldlines \cite{Rovelli:1990jm}-\cite{Rovelli:1995}. For clock time to become able to provide information about the local parametric time of each single worldline, a condition is required on the value of the component $u^0$ of (\ref{partlight03}) relating clock time and parametric time. Only when this relationship is established, one can address the more difficult non-local problem inherent in the fact that, in principle, a clock evolving along a worldline A cannot measure time defined on a different worldline B. Synchronization of distant events can be performed in at least two different ways, invoking absolute and relative simultaneity respectively, which we study separately.

\section{Absolute simultaneity}

First we consider pre-relativistic global time giving rise to a spacetime structure which presents itself as independent from observers, and thus as {\it absolute}. In order to construct such geometry, we introduce a preferred parametric time direction in spacetime, and we impose a common rate of change $u^0$ of all locally measurable clock times with respect to parametric time $\tau$, so that also $(w\rfloor d\tau )$ becomes fixed.

\subsection{Absolute space and time}

We postulate absolute time to be associated to a preferred congruence of curves with tangent vector ${\buildrel abs\over{u}}$, such that ${\buildrel abs\over{u}}\rfloor d{\buildrel abs\over{\tau}} =1$ when contracted with the parametric time differential $d{\buildrel abs\over{\tau}}$ defining a preferred orientation on $T^*(\Sigma )$. (In the following, we denote these quantities simply as $u$ and $d\tau$ respectively.) Next we choose the time component of the tetrad (\ref{partlight09}) to be aligned with the unique parametric time direction $d\tau$. That is, we impose the {\it time gauge} condition
\begin{equation}
\underline{\vartheta}^0 =0\,,\label{absoltime01}
\end{equation}
so that from (\ref{partlight09}) follows
\begin{equation}
\vartheta ^0 = d\tau\,u^0\,.\label{absoltime02}
\end{equation}
Due to the fact that tetrads are (nonlinear) translational connections, (\ref{absoltime02}) plays the role of a {\it time connection} defining horizontal slices orthogonal to it, expanded by the spatial basis vectors $\{ e_a\}$ such that $e_a\rfloor\vartheta ^0 =0$, implying $e_a\rfloor d\tau =0$. Since (\ref{absoltime02}) satisfies the Frobenius foliation condition $\vartheta ^0\wedge d\vartheta ^0 =0$, the spatial hypersurfaces coincide with the simultaneity slices of the spacetime foliation along $u$.

Now we consider curves oblique to $u$ (that is, at relative motion with respect to $u$), with tangent vector $w$, whose components we have found to have the general form (\ref{partlight07}). Definition (\ref{partlight06}) with (\ref{absoltime01}) implies that $v^0 =0$, so that the zero component of (\ref{partlight07}) reads
\begin{equation}
w^0 = (w\rfloor d\tau )\,u^0\,.\label{absoltime03}
\end{equation}
From (\ref{partlight03}) we know that $u^0$ measures the rate of change of the clock time variable $\xi ^0$ with respect to parametric time as
\begin{equation}
u^0 = {\cal \L\/}_u\xi ^0 +{\buildrel (T)\over{\Gamma _{\bot}^0}}\,.\label{light07bis}
\end{equation}
A postulate on the value of $u^0$ (for instance $u^0 =\kappa$, with $\kappa$ as a constant) is required to establish a yet not existing relationship between phenomenological clock time and theoretical parametric time. However, for the moment we do not fix $u^0$, but we assume such time rate to be the same for all observers, no matter if they are at rest or moving in space, that is
\begin{equation}
w^0 = u^0\,.\label{absoltime04}
\end{equation}
Eq.(\ref{absoltime03}) with (\ref{absoltime04}) then yields
\begin{equation}
(w\rfloor d\tau )=1\,,\label{absoltime05}
\end{equation}
which provides us with a value of $(w\rfloor d\tau )$, as we were looking for. By replacing (\ref{absoltime05}) in (\ref{partlight07}), one gets the pre-relativistic composition of velocities
\begin{equation}
w^a = u^a + v^a\,.\label{absoltime06}
\end{equation}
The vectors $u$ and $w$, when describing trajectories of particles at absolute rest (independently of if it is observationally possible to determine if this is the case), are orthogonal to $e_a$. Otherwise, they are oblique or curved with respect to the simultaneity slices expanded by the spatial basis vectors, having a common component $w^0=u^0$.

Absolute space and time are conceived as owning an intrinsic structure independent of observers, where a preferred global frame (\ref{5dpoinc72b}) is defined. However, actually, only the direction of time is universally fixed. Some symmetries of space making impossible to determine absolute spatial orientations or absolute point localizations were pointed out already in Greek Antiquity. For instance, the universe models of Anaximander and Aristotle possessed rotational symmetry. They had an absolute center coinciding with that of the Earth, but the primitive assumption of absolute up and down was disregarded in favor of the physical indistinguishability of directions pointing to the center of the universe. On the other hand, an infinite homogeneous space with more or less well defined translational symmetry was postulated by the Atomists. In general, the metaphysical conception of absolute space and time can be maintained if desired even if one admits the direct experience of them to be restricted by Euclidean and time translational symmetry \cite{Friedman} or by the larger symmetry to be studied next.

\subsection{Galilei spacetime}

Let us go a step further by requiring the compatibility of the former general approach to absolute time with Newtonian mechanics. Then one has to consider the Galilei group of transformations, whose finite form is presented in Appendix C, preserving both, universality of time (invariance of time rate (\ref{galvel02}), of duration (\ref{galitetrad02}), etc.), and the validity of Newton's laws of motion.

Galilean geometry can be endowed with a metric structure assigning to an arbitrary four-vector $X$, on the one hand, its invariant time component $X^0 := X\rfloor\vartheta ^0$ (see (\ref{galvel02}), (\ref{galaccel02})), etc.), and on the other hand its spatial norm, defined only for 3-dimensional hypersurfaces with $X^0 =0$ as the squared length $|X|^2 =\delta _{ab}\,X^a X^b$ built with the Euclidean metric. See \cite{Weyl}, pg.156. Absolute time duration (\ref{galitetrad02}) as much as the Euclidean metric defined in each simultaneity hypersurface are Galilei invariants, being the length of a segment the invariant distance between its simultaneous extreme points.

Regarding the postulate needed to relate clock time and parametric time so that the latter becomes measurable by a clock, let us impose the condition
\begin{equation}
{\cal \L\/}_u u^0 =0\,,\label{galaccel04}
\end{equation}
invariant according to (\ref{galaccel02}), so that (\ref{galaccel03}) yields
\begin{equation}
\hat{{\cal \L\/}}_u \hat{u}^a = (\mathfrak{R}^{-1})_b{}^a \,{\cal \L\/}_u u^b \,,\label{galaccel05}
\end{equation}
showing that the acceleration (but not the velocity, see (\ref{galvel03})) transforms as a vector in the three-dimensional space. In the simplified case of absence of connections, the solution of (\ref{galaccel04}) is $u^0 =\kappa$, being $\kappa$ a constant, so that (\ref{light07bis}) reduces to $u^0 = {\it l}_u\,\xi ^0 =\kappa$, implying $\xi ^0 = \kappa\tau + const.$, thus allowing clocks to measure absolute time. The constant $\kappa$ (with dimensions of velocity but having not to do with the speed of light) doesn't play any essential role. It can be absorbed in the relevant physical fields by redefining them as as $t:=\xi ^0/\kappa$, $\tilde{v}^a := \kappa\beta ^a$, $e_t:= \kappa e_0$, etc., so that, for instance, $t=\tau + const.$, (\ref{galframe01}) takes the form $\hat{e}_t = e_t + \tilde{v}^a e_a$, etc.

According to the Galilei principle of relativity, all reference frames related by Galilei transformations are equivalent regarding the description of classical mechanics, since Newton's laws of motion are preserved by the Galilei group. In particular, the law of inertia (that is, of vanishing acceleration of free bodies) is Galilei-invariant due to (\ref{galaccel05}). Accordingly, even if absolute space and time do exist, no dynamical effects allow to distinguish absolute rectilinear motion (for instance that of the Earth in space) from absolute rest.

Let us relate the Galilei transformations to the results of previous section on relative velocities in spacetime with absolute simultaneity. We write (\ref{absoltime04}) and (\ref{absoltime06}) as
\begin{equation}
w = u + v\,,\label{glob03}
\end{equation}
and we evaluate $u$ in its rest frame with $\hat{u}^a=0$, where it reduces to
\begin{equation}
u = u^\alpha e_\alpha = \hat{u}^0 \hat{e}_0\,.\label{glob01}
\end{equation}
Taking into account (\ref{galframe01}) and (\ref{galvel02}), one can express (\ref{glob01}) as
\begin{equation}
u = u^0\,\left(\,e_0 + \beta ^a e_a\,\right)\,.\label{glob02}
\end{equation}
On the other hand, in the same frame where $\hat{u}^a=0$, the velocity (\ref{glob03}) takes the form
\begin{equation}
w = \hat{u}^0 \hat{e}_0 +\hat{v}^a \hat{e}_a\,.\label{glob04}
\end{equation}
Being $\hat{u}^0=u^0$ according to (\ref{galvel02}), if we identify the spatial components of the transversal velocity (\ref{partlight06}) with the group parameters $\beta ^a$ (times $u^0$) as
\begin{equation}
\hat{v}^a = u^0 \beta ^a\,,\label{glob05}
\end{equation}
Eq.(\ref{glob04}) can be rewritten as
\begin{equation}
w = u^0\,\left(\,\hat{e}_0 + \beta ^a \hat{e}_a\,\right)\,.\label{glob06}
\end{equation}
Thus (\ref{glob06}), with the same components as (\ref{glob02}) but referred to a different frame, can be regarded as the result of an active Galilei transformation of (\ref{glob02}) with boost parameter proportional to the relative transversal velocity. Expressing $u$ and $w$ in the rest frame of $u$, see (\ref{glob01}), this transformation acts as
\begin{equation}
u = \hat{u}^0\,\hat{e}_0\longrightarrow  w = u^0\,\left(\,\hat{e}_0 + \beta ^a \hat{e}_a\,\right)\,,\label{activeGal}
\end{equation}
mapping the fourvelocity $u$ of an observer {\it at rest} to the fourvelocity $w$ of a body moving along an oblique worldline. Application (\ref{activeGal}) constitutes a particular case of the general active Galilei transformations
\begin{equation}
u = u^\alpha e_\alpha =\hat{u}^\alpha\hat{e}_\alpha\longrightarrow  w = u^\alpha\hat{e}_\alpha\label{activeGalbis}
\end{equation}
of fourvelocity vectors.

Let us finally comment that, given the position (\ref{5dpoinc73}) referred to an orthogonal frame as $p = \mathfrak{o} +\xi ^\alpha\,e_\alpha$, it can be passively Galilei-transformed as $p = \hat{\mathfrak{o}} +\hat{\xi}^\alpha\,\hat{e}_\alpha\,$, in such a way that the time piece, found from (\ref{galframe01}) and (\ref{galcoord01}) to be $\hat{\xi}^0 \hat{e}_0 = (\xi ^0 -a^0)( e_0 +\beta ^a e_a )$, becomes oblique with respect to $\xi ^0 e_0$, while the spatial part $\hat{\xi}^a \hat{e}_a = [\,\xi ^a - a^a -\beta ^a(\xi ^0 - a^0 ) ]\,e_a$ (see (\ref{galframe02}) and (\ref{galcoord02})) keeps the original orientation of $\xi ^a e_a$. No inclination of the simultaneity hypersurfaces occur. They remain {\it horizontal} for all observers, guaranteeing absolute simultaneity.

\section{Relative simultaneity}

In the previous section, absolute time and absolute simultaneity were derived from the postulate of the existence of worldlines with a preferred time vector ${\buildrel abs\over{u}}$ defining absolute time orientation, with an universal zero component $u^0$ (subjected to the condition ${\cal \L\/}_u u^0 =0$) ensuring the time rate to be common to all observers, so that measurable clock time $\xi^0$ relates to (absolute) parametric time in the same simple way in all reference frames.

Relative simultaneity to be introduced next also requires preferred worldlines to exist, namely those of light rays, oblique to any time vector $u$, whose tangent vector $w_L$ defines light cones with {\it absolute} (that is, invariant) spacetime orientation. The time rate will be fixed to be $\hat{u}^0 =c$ (equal to the constant speed of light) for any observer at rest, thus allowing clock time to measure parametric time as proper time, while in general the time rate $u^0$ calculated for moving bodies is found to depend on relative velocities. Einstein's synchronization postulate provides a guide to determine the Minkowski metric of spacetime.

\subsection{Construction of a metric space}

The structuring assumptions to be introduced in the following, concerning relative simultaneity, are of metric nature. So, first we have to extend our original affine framework to a metric-affine geometry by including a metric \cite{Hehl:1995ue}, and then we determine the form of the latter in view of suitable hypotheses.

We define a (pseudo-Riemannian) metric tensor \cite{Nakahara} as a map $g: T_p(\Sigma)\times T_p(\Sigma)\rightarrow \mathbb{R}$ such that
\begin{eqnarray}
&& g(X\,,Y)=g(Y\,,X)\,,\label{metric01}\\
&& g(X\,,Y)=0\hskip0.2cm\text{for all}\hskip0.15cm Y \iff X=0\,,\label{metric02}\\
&& g(X\,,a\,Y +b\,Z)= a\,g(X\,,Y) + b\,g(X\,,Z)\,.\label{metric03}
\end{eqnarray}
When applied to the frame vectors $\left\{ e_\alpha\right\}$, it yields the metric tensor components
\begin{equation}
g_{\alpha\beta} := g(e_\alpha\,,e_\beta )\,,\label{orthog01}
\end{equation}
symmetric by definition, but with otherwise general values for its components, as derived from mutually oblique basis vectors. We use (\ref{orthog01}) for raising and lowering indices so that the standard notation $g_{\alpha\beta}X^\alpha Y^\beta = X_\alpha Y^\alpha$ holds.

In terms of the four-velocity (\ref{partlight03}) and taking (\ref{orthog01}) into account, we define the projector
\begin{equation}
h_\beta{}^\alpha :=\delta _\beta ^\alpha - {{u_\beta\,u^\alpha}\over{(u_\mu u^\mu )}}\,,\label{partlight10}
\end{equation}
allowing one to write down the identity
\begin{equation}
\vartheta ^\alpha \equiv \vartheta ^\beta\left[ {{u_\beta\,u^\alpha}\over{(u_\mu u^\mu )}} + h_\beta{}^\alpha\right]\,.\label{partlight11}
\end{equation}
Comparing (\ref{partlight09}) with (\ref{partlight11}) we identify
\begin{equation}
d\tau = {{u_\beta\,\vartheta ^\beta}\over{(u_\mu u^\mu )}}\,,\label{partlight12}
\end{equation}
and
\begin{equation}
\underline{\vartheta}^\alpha = \vartheta ^\beta\,h_\beta{}^\alpha\,,\label{partlight13}
\end{equation}
so that from (\ref{partlight13}) with (\ref{partlight10}) follows
\begin{equation}
u_\alpha\,\underline{\vartheta}^\alpha = 0\,,\label{partlight14}
\end{equation}
yielding, together with definition (\ref{partlight06}), the orthogonality condition
\begin{equation}
u_\alpha\,v^\alpha = 0\,.\label{partlight15}
\end{equation}
Consequently, the squared norm of velocities (\ref{partlight07}) oblique to $u$ reads
\begin{equation}
w_\alpha w^\alpha = (w\rfloor d\tau )^2\,\left(\,u_\alpha u^\alpha + v_\alpha v^\alpha\,\right)\,.\label{partlight16}
\end{equation}
Among all worldlines with tangent vectors satisfying (\ref{partlight16}), light signals are singled out by the postulate of constancy of light speed (plus an additional condition on the time rate, as we will see), having as a consequence the complete determination of the metric, and thus making possible to fix $(w\rfloor d\tau )$ for particle trajectories.

\subsection{From oblique to orthogonal basis vectors}

Let us simplify things by evaluating the metric (\ref{orthog01}) referred to a suitable vector basis. Actually, without loss of generality, one can introduce a local frame $\left\{\tilde{e}_\alpha\right\}$ choosing its time vector component to be orthogonal to the spatial ones. To do so, we make use of suitable lapse and shift functions $N$ and $N^a$ allowing to express the old vectors in terms of the new ones as
\begin{eqnarray}
e_0 &=& N \tilde{e}_0 + N^a \tilde{e}_a\,,\label{orthog02}\\
e_a &=& \tilde{e}_a\,,\label{orthog03}
\end{eqnarray}
so that the new frame vectors read
\begin{eqnarray}
\tilde{e}_0 &=& {1\over N}\left(\,e_0 - N^a e_a\right)\,,\label{orthog04}\\
\tilde{e}_a &=& e_a \,.\label{orthog05}
\end{eqnarray}
In terms of them we define the metric
\begin{equation}
\tilde{g}_{\alpha\beta} = g(\tilde{e}_\alpha\,,\tilde{e}_\beta )\,,\label{orthog06}
\end{equation}
whose components, in view of (\ref{metric03}), relate to those of (\ref{orthog01}) as
\begin{eqnarray}
\tilde{g}_{_{00}} &=& {1\over{N^2}}\left(\,g_{_{00}} - 2\,g_{a0}N^a + g_{ab}N^a N^b\right)\,,\label{orthog07}\\
\tilde{g}_{a0} &=& {1\over N}\left( g_{a0}-g_{ab} N^b\right)\,,\label{orthog08}\\
\tilde{g}_{ab} &=& g_{ab}\,.\label{orthog09}
\end{eqnarray}
The orthogonality between $\tilde{e}_a$ and $\tilde{e}_0$ requires (\ref{orthog08}) to vanish. Thus we impose the condition
\begin{equation}
g_{a0}-g_{ab} N^b =0\,,\label{orthog10}
\end{equation}
defining the shift functions $N^b$ in terms of metric tensor components, so that (\ref{orthog07})-(\ref{orthog09}) reduce to
\begin{eqnarray}
\tilde{g}_{_{00}} &=& {1\over{N^2}}\left(\,g_{_{00}} - g_{ab}N^a N^b\right)\,,\label{orthog11}\\
\tilde{g}_{a0} &=& 0\,,\label{orthog12}\\
\tilde{g}_{ab} &=& g_{ab}\,.\label{orthog13}
\end{eqnarray}
In the new basis, the components of an arbitrary vector, say $w= w^\alpha e_\alpha$, relate to the old ones as
\begin{eqnarray}
\tilde{w}^0 &=& N w^0\,,\label{orthog14}\\
\tilde{w}^a &=& w^a + w^0 N^a\,,\label{orthog15}
\end{eqnarray}
in such a way that
\begin{equation}
w=\tilde{w}^\alpha \tilde{e}_\alpha = w^\alpha e_\alpha\,,\label{orthog16}
\end{equation}
and
\begin{eqnarray}
w_\alpha w^\alpha &:=& g_{\alpha\beta} w^\alpha w^\beta\nonumber\\
&=&\tilde{g}_{\alpha\beta} \tilde{w}^\alpha \tilde{w}^\beta\nonumber\\
&=& \tilde{g}_{_{00}} \tilde{w}^0 \tilde{w}^0 +\tilde{g}_{ab} \tilde{w}^a \tilde{w}^b\,.\label{orthog17}
\end{eqnarray}
Moreover, in an analogous way, it is possible to choose mutually orthogonal spatial vectors as follows. Let us multiply them by a certain matrix $\sqrt{k}\,z_a{}^b$ so that (\ref{orthog04}) and (\ref{orthog05}) transform into
\begin{eqnarray}
\tilde{\tilde{e}}_0 &=& \tilde{e}_0 \,,\label{orthog18}\\
\tilde{\tilde{e}}_a &=& \sqrt{k}\,z_a{}^b \tilde{e}_b \,,\label{orthog19}
\end{eqnarray}
with $k$ as a quantity to be fixed. The time metric tensor contribution due to (\ref{orthog18}) coincides with (\ref{orthog11}), that is
\begin{equation}
\tilde{\tilde{g}}_{_{00}} =\tilde{g}_{_{00}}\,,\label{orthog20}
\end{equation}
while the spatial metric tensor built from (\ref{orthog19}) reads
\begin{equation}
\tilde{\tilde{g}}_{ab} = k\,z_a{}^c z_b{}^d \tilde{g}_{cd}\,.\label{orthog21}
\end{equation}
In analogy with (\ref{orthog10}), we impose in (\ref{orthog21}) the condition
\begin{equation}
z_a{}^c z_b{}^d \tilde{g}_{cd} = \delta _{ab}\,,\label{orthog22}
\end{equation}
defining the $z_a{}^b$ matrices in terms of the metric tensor components $\tilde{g}_{cd}$, so that (\ref{orthog21}) simplifies to
\begin{equation}
\tilde{\tilde{g}}_{ab} = k\,\delta _{ab}\,.\label{orthog22}
\end{equation}
The vector components of a vector $w= w^\alpha e_\alpha$ in the basis (\ref{orthog18}), (\ref{orthog19}) are
\begin{eqnarray}
\tilde{\tilde{w}}^0 &=& \tilde{w}^0 = N w^0\,,\label{orthog23}\\
\tilde{\tilde{w}}^a &=& \tilde{w}^b (z^{-1})_b{}^a = (w^b + w^0 N^b)(z^{-1})_b{}^a\,,\label{orthog24}
\end{eqnarray}
in terms of which (\ref{orthog16}) becomes extended to
\begin{equation}
w=\tilde{\tilde{w}}^\alpha \tilde{\tilde{e}}_\alpha = \tilde{w}^\alpha \tilde{e}_\alpha = w^\alpha e_\alpha\,,\label{orthog16bis}
\end{equation}
and (\ref{orthog17}) can be expressed with the help of (\ref{orthog20}) and (\ref{orthog22}) as
\begin{eqnarray}
w_\alpha w^\alpha &:=& g_{\alpha\beta} w^\alpha w^\beta\nonumber\\
&=&\tilde{\tilde{g}}_{\alpha\beta} \tilde{\tilde{w}}^\alpha \tilde{\tilde{w}}^\beta\nonumber\\
&=& \tilde{\tilde{g}}_{_{00}} \tilde{\tilde{w}}^0 \tilde{\tilde{w}}^0 + k\,\delta _{ab} \tilde{\tilde{w}}^a \tilde{\tilde{w}}^b\,.\label{orthog17bis}
\end{eqnarray}
Besides the value of $k$, the only metric tensor element still remaining undetermined is $\tilde{\tilde{g}}_{_{00}}$. We will fix it making use of Einstein's synchronization criterion.

\subsection{Light signals, Einstein's synchronization and Minkowski metric}

 We know that a clock placed at an arbitrary point measures local clock time. Thus, although it can certainly determine the round-trip time of a reflected signal by measuring its times of departure and arrival, it is not possible for it to measure time intervals whose initial and final instants occur at separate points of space. In order to synchronize distant events, Einstein stipulates that the one-way speed of light coincides with the mean velocity of light, when measured in a round-trip, which is known to be a constant $c$. Light is supposed to propagate with equal velocity $c$ in all directions relative to a given congruence of curves. This means that, in the general expression (\ref{partlight07}) for oblique vectors, when dealing with light worldlines one has to take the transversal velocity to be $v^\alpha = c\,n^\alpha$, with $n^\alpha$ as a unit vector. (The possible difficulty in interpreting the zero component of $n^\alpha$ is obviated by the fact that it vanishes in the particular frame we are going to consider immediately.) The light tangent vector components (\ref{partlight07}) thus read
\begin{equation}
w_{_L}^\alpha = (w_{_L}\rfloor d\tau )\,\left(\,u^\alpha + c\,n^\alpha\,\right)\,,\label{light01}
\end{equation}
with squared norm
\begin{equation}
w^{_L}_\alpha w_{_L}^\alpha = (w_{_L}\rfloor d\tau )^2\,\left(\,u_\alpha u^\alpha + c^2\,n_\alpha n^\alpha\,\right)\label{light02}
\end{equation}
as a particular case of (\ref{partlight16}). We notice that the standard result of vanishing (\ref{light02}) does not follow automatically. In order to get it, the synchronization convention by means of light signals demands additional assumptions. Ignoring other options, we are going to impose (\ref{light02}) to be null by introducing suitable constructive hypotheses, while the choice of alternative postulates giving rise to different geometries remains an open possibility. Vanishing light (squared) norm (\ref{light02}), for $(w_{_L}\rfloor d\tau )\neq 0$ but otherwise undetermined, requires
\begin{equation}
u_\alpha u^\alpha + c^2\,n_\alpha n^\alpha =0\,.\label{light03}
\end{equation}
Let us evaluate (\ref{light03}) in the particular frame where the components of $u$ reduce to $\hat{u}^a=0$, $\hat{u}^0\neq 0$. According to (\ref{partlight15}) with the orthogonality conditions (\ref{orthog12}) and (\ref{orthog22}), in such frame $\hat{n}^0 =0$, so that (\ref{light03}) reduces to
\begin{equation}
\hat{g}_{_{00}}(\hat{u}^0)^2 + c^2\,k\,\hat{n}_a \hat{n}^a =0\,,\label{light04}
\end{equation}
where $\hat{n}^a$ is to be interpreted as the unit vector normal to the wave front, being $\hat{n}_a \hat{n}^a =1$ since we took it to be a unit vector. Eq.(\ref{light04}) forces one to take either $\hat{g}_{_{00}}< 0$ and $k > 0$ or $\hat{g}_{_{00}} >0$ and $k <  0$. One could take $k$ to be a -non necessarily constant- conformal factor. However, disregarding this possibility, we choose $k=+1$, and we postulate the condition $\hat{g}_{_{00}}(\hat{u}^0)^2 =-c^2$ to be fulfilled by the separate condition
\begin{equation}
\hat{u}^0 =c\,,\label{light05}
\end{equation}
and the choice
\begin{equation}
\hat{g}_{_{00}}=-1\label{light06}
\end{equation}
for the previously undetermined metric component. Eq.(\ref{light05}) plays a role analogous to (\ref{galaccel04}). Indeed, in view of (\ref{partlight03}), it yields
\begin{equation}
c =\hat{u}^0 = \hat{\cal \L\/}_u\hat{\xi}^0 +{\buildrel (T)\over{\hat{\Gamma}_{\bot}^0}}\,,\label{light07}
\end{equation}
thus fixing the rate of change of clock time with respect to parametric time to be a constant when measured by an observer at rest. So we get a (conventional) relation between a directly measurable and an in principle non observable quantity, as in the absolute time case of Section IV. In the special-relativistic limit where connections can be put equal to zero, (\ref{light07}) implies that, for observers at rest, $\hat{\xi}^0 =c\tau + const$, so that parametric time becomes measurable as proper time by a local clock.

On the other hand, the result (\ref{light06}) can be put together with (\ref{orthog12}) and (\ref{orthog22}) (with $k=+1$), at least in the particular frame considered so far, to complete the Minkowski metric
\begin{equation}
\hat{g}_{\alpha\beta}=o_{\alpha\beta}:=diag(-+++)\,.\label{Mink}
\end{equation}
In that frame, (\ref{Mink}) can be used to write (\ref{light03}) as
\begin{equation}
u_\alpha u^\alpha =o_{\alpha\beta}\,u^\alpha u^\beta =-c^2\,,\label{lightnorm}
\end{equation}
assigning a fixed norm to the time tangent vector $u$ (for which only its direction was determined until now). In principle, the Minkowski metric (\ref{Mink}) is referred to a particular convergence of curves of observers at relative rest, constituting an {\it extended reference frame} sharing a common parametric time. In order for the synchronization criterion to become valid for observers at relative motion, the value of $c$, {\it constant} for rest observers, is required to remain {\it invariant} under suitable transformations involving relative velocities.

\subsection{Invariance of the Minkowski metric}

The postulate of invariance of the light speed $c$, consistent with the validity of the Maxwell equations in relatively moving frames, means that (\ref{lightnorm}) has to transform as $o_{\alpha\beta}\,u^\alpha u^\beta = o_{\alpha\beta}\,\hat{u}^\alpha \hat{u}^\beta =-c^2$, or equivalently, that the Minkowski metric (\ref{Mink}) must be left invariant as
\begin{equation}
o_{\alpha\beta} = \Lambda _\alpha{}^\mu \Lambda _\beta{}^\nu  o_{\mu\nu}\,.\label{Lorcondit01}
\end{equation}
The $GL(4\,,\mathbb{R})$ transformation matrices $\Lambda _\alpha{}^\beta$ present in (\ref{posit01})-(\ref{posit03}), restricted by condition (\ref{Lorcondit01}), are the Lorentz matrices. (For clarity, we consider finite matrices. They are such that $\Lambda _\alpha{}^\beta =(\Lambda ^{-1})^\beta{}_\alpha\,$.) The extension of the synchronization criterion to relatively moving systems has as a consequence that the measurements performed in different frames are related  to each other through Poincar\'e transformations acting as (\ref{posit01})-(\ref{posit03}), in such a way that quantities which are not directly observable become calculable.

The Lorentz matrices decompose into the product of rotations times boosts as in (\ref{Lormatrix}), with both, the boost matrix $\mathfrak{B}_\alpha{}^\beta$ and the rotation matrix $\mathfrak{R}_\alpha{}^\beta$, obeying (\ref{Lorcondit01}) separately as
\begin{equation}
o_{\alpha\beta} = \mathfrak{B}_\alpha{}^\mu \mathfrak{B}_\beta{}^\nu  o_{\mu\nu}\,,\label{Lorcondit03}
\end{equation}
and an analogous equation for $\mathfrak{R}_\alpha{}^\beta$. The rotational matrix is the same as given in (\ref{rotmatrix01}) and (\ref{rotmatrix02}). On the other hand, in order to find the (finite) boost matrix, we solve (\ref{Lorcondit03}) expanding it in components as
\begin{eqnarray}
-1 &=& -(\mathfrak{B}_0{}^0)^2 + \mathfrak{B}_0{}^a \mathfrak{B}_0{}^b \delta _{ab}\,,\label{boostcond01}\\
0 &=& -\mathfrak{B}_a{}^0 \mathfrak{B}_0{}^0 + \mathfrak{B}_a{}^b \mathfrak{B}_0{}^c \delta _{bc}\,,\label{boostcond02}\\
\delta _{ab} &=& -\mathfrak{B}_a{}^0 \mathfrak{B}_b{}^0 +\mathfrak{B}_a{}^c \mathfrak{B}_b{}^d \delta _{cd}\,,\label{boostcond03}
\end{eqnarray}
and introducing the notation
\begin{equation}
\gamma := \mathfrak{B}_0{}^0\,,\quad \gamma\beta ^a :=\mathfrak{B}_0{}^a\,,\label{gammabetadefs}
\end{equation}
being $\gamma$ and $\beta ^a$ not constant in general. Replacing (\ref{gammabetadefs}) in (\ref{boostcond01})-(\ref{boostcond03}) and suitably combining
equations (\ref{boostcond02}) and (\ref{boostcond03}), we get the conditions
\begin{eqnarray}
\gamma ^2 &=& {1\over{(\,1-\beta ^2)}}\,,\label{boostcond04}\\
\mathfrak{B}_a{}^0 &=& \mathfrak{B}_a{}^b \beta _b\,,\label{boostcond05}\\
\delta _{ab} &=& \mathfrak{B}_a{}^c \mathfrak{B}_b{}^d (\,\delta _{cd}- \beta _c\beta _d )\,.\label{boostcond06}
\end{eqnarray}
With (\ref{boostcond04}) expressing $\gamma$ in terms of $\beta ^a$, solutions of (\ref{boostcond05}) and (\ref{boostcond06}), together with definitions (\ref{gammabetadefs}), yield the complete set of boost matrix components
\begin{eqnarray}
\mathfrak{B}_0{}^0 &=&\gamma\,,\hskip0.7cm \mathfrak{B}_0{}^b = \gamma\beta ^b\label{boost03}\\
\mathfrak{B}_a{}^0 &=&\gamma\beta _a\,,\quad \mathfrak{B}_a{}^b = \delta _a^b +(\gamma -1){{\beta _a\beta ^b}\over{\beta ^2}}\,,\label{boost04}
\end{eqnarray}
where
\begin{equation}
\gamma :={1\over{\sqrt{1-\beta ^2}}}\,.\label{gamma}
\end{equation}
The physical meaning of the group parameters $\beta ^a$ (and of $\gamma$) as velocity-shaped quantities is illustrated by considering the transformation of the fourvelocity components (\ref{partlight03}), that is
\begin{equation}
\hat{u}^\alpha = ( \mathfrak{B}^{-1})_\beta{}^\alpha\,u^\beta\,,\label{boostedu01bis}
\end{equation}
whose inverse relation
\begin{equation}
u^\alpha = \mathfrak{B}_\beta{}^\alpha\,\hat{u}^\beta\,,\label{boostedu02bis}
\end{equation}
expressed in terms of (\ref{boost03}) and (\ref{boost04}), reads
\begin{eqnarray}
u^0 &=& \gamma\left(\,\hat{u}^0 + \beta _a \hat{u}^a\,\right)\,,\label{hatu0bisbis}\\
u^a &=& \hat{u}^a +(\gamma -1){{\beta ^a\beta _b}\over{\beta ^2}}\,\hat{u}^b + \gamma\beta ^a \hat{u}^0\,.\label{hatuabisbis}
\end{eqnarray}
The rest frame conditions $\hat{u}^a =0$, $\hat{u}^0 =c$ (see (\ref{light05})), when replaced in (\ref{hatu0bisbis}) and (\ref{hatuabisbis}) yield
\begin{eqnarray}
u^0 &=& c\,\gamma\,,\label{u0bis}\\
u^a &=& c\,\gamma\beta ^a\,,\label{uabis}
\end{eqnarray}
showing a direct relationship between the Lorentz boost parameters and the fourvelocity (\ref{partlight03}) (expressed in terms of bundle fields). Notice that (\ref{u0bis}) differs from (\ref{light05}) in that the time rate of a moving system is velocity-dependent.

According to (\ref{posit01}), boosts (\ref{boost03}), (\ref{boost04}) act on frame vectors as
\begin{equation}
\hat{e}_\alpha = \mathfrak{B}_\alpha{}^\beta e_\beta\,,\label{boost01}
\end{equation}
and on tetrads (covector-valued 1-forms) as
\begin{equation}
\hat{\vartheta}^\alpha = ( \mathfrak{B}^{-1})_\beta{}^\alpha\,\vartheta ^\beta = \mathfrak{B}^\alpha{}_\beta\,\vartheta ^\beta\,,\label{boost02}
\end{equation}
that is
\begin{eqnarray}
\hat{e}_0 &=& \gamma\left(\,e_0 +\beta ^a e_a\,\right)\,,\label{hate0}\\
\hat{e}_a &=& e_a +(\gamma -1){{\beta _a\beta ^b}\over{\beta ^2}}\,e_b +\gamma\beta _a e_0\,.\label{hatea}
\end{eqnarray}
and
\begin{eqnarray}
\hat{\vartheta}^0 &=& \gamma\left(\,\vartheta ^0 - \beta _a\vartheta ^a\,\right)\,,\label{hattheta0}\\
\hat{\vartheta}^a &=& \vartheta ^a +(\gamma -1){{\beta ^a\beta _b}\over{\beta ^2}}\,\vartheta ^b -\gamma\beta ^a\vartheta ^0\,.\label{hatthetaa}
\end{eqnarray}
The invariance of the Minkowski metric, together with transformations (\ref{boost02}), implies the invariance of the line element
\begin{equation}
ds^2 = o_{\alpha\beta}\vartheta ^\alpha\otimes\vartheta ^\beta\,,\label{lineelement}
\end{equation}
so that the congruence of distant intervals is guaranteed. Analogously, congruent vectors with equal norm are also defined at arbitrary points.

The composition of boosts (\ref{boost03}), (\ref{boost04}), having the same form (\ref{velcomp01}) as in the Galilei case, yields the relativistic velocity composition law
\begin{equation}
b^a = {{\hat{b}^a +\beta ^a}\over{(1+\hat{b}_c\beta ^c)}} +{{(\,1/\gamma -1\,)\,\hat{b}^b}\over{(1+\hat{b}_c\beta ^c)}}\left( \delta _b^a -{{\beta _b\beta ^a}\over{\beta ^2}}\right)\,,\label{velcomp02}
\end{equation}
quite different from (\ref{galcomp}).

\subsection{Relation between relativistic particle trajectories}

Let us return to Eq.(\ref{partlight07}) relating the velocities of systems at relative motion and consider the case of trajectories other tan those of light rays, that is, with $w_\alpha w^\alpha\neq 0$. For later convenience, we rewrite $v^\alpha$ in (\ref{partlight06}) as $v^\alpha = c\,b^\alpha$ so that (\ref{partlight16}) reads
\begin{equation}
w_\alpha w^\alpha = (w\rfloor d\tau )^2\,\left(\,u_\alpha u^\alpha + c^2 b_\alpha b^\alpha\,\right)\,.\label{sqrdnorm}
\end{equation}
Instead of the condition (\ref{absoltime04}) imposed in the context of absolute time, assigning a common zero component (universal time rate) to $w$ and $u$, now we require $w$ and $u$, defined along different trajectories, to be congruent tangent vectors, that is, such that
\begin{equation}
w_\alpha w^\alpha = u_\alpha u^\alpha =-c^2\,,\label{congruence01}
\end{equation}
in accordance with (\ref{lightnorm}). From (\ref{sqrdnorm}) and (\ref{congruence01}), the until now undetermined quantity $(w\rfloor d\tau )$ (equal to $-{1\over c^2}\,u_\alpha w^\alpha$, as deduced from (\ref{partlight07}) with (\ref{partlight15}) and (\ref{lightnorm})) is found to be
\begin{equation}
(w\rfloor d\tau ) ={1\over{\sqrt{1-b_\beta b^\beta}}}\,,\label{partlight23}
\end{equation}
(compare with (\ref{absoltime05})), and replacing (\ref{partlight23}) in (\ref{partlight07}) one gets
\begin{equation}
w^\alpha = {1\over{\sqrt{1-b_\beta b^\beta}}}\,\left(\,u^\alpha + c\,b^\alpha\,\right)\,.\label{partlight17}
\end{equation}
In the rest frame of $u$, where $\hat{u}^a=0$ and (\ref{light05}) holds, and where (\ref{partlight15}) reduces to $u_\alpha\,v^\alpha = u_\alpha\,c\,b^\alpha =c\,\hat{u}_0\,\hat{b}^0 = -c^2\,\hat{b}^0 =0$, so that $\hat{b}^0 =0$, Eq.(\ref{partlight17}) gives rise to the following chain of equalities
\begin{eqnarray}
w = w^\alpha e_\alpha &=& {1\over{\sqrt{1-b_\beta b^\beta}}}\,\left(\,u^\alpha + c\,b^\alpha\,\right)e_\alpha \nonumber\\
&=& {1\over{\sqrt{1-\hat{b}_c \hat{b}^c}}}\,\left(\,\hat{u}^0\hat{e}_0 + c\,\hat{b}^a \hat{e}_a\,\right)\nonumber\\
&=& {c\over{\sqrt{1-\hat{b}^2}}}\,\left(\,\hat{e}_0 + \hat{b}^a \hat{e}_a\,\right)\,.\label{partlight18}
\end{eqnarray}
Let us compare the final expression of (\ref{partlight18}) with that of $u$ in the same frame with $\hat{u}^a=0$, that is $u = \hat{u}^0 \hat{e}_0$. Replacing the values of $\hat{u}^0$ and $\hat{e}_0$ as given by (\ref{light05}) and (\ref{hate0}) respectively, one finds
\begin{equation}
u = u^\alpha e_\alpha = \hat{u}^0 \hat{e}_0 = c\gamma\,\left(\,e_0 + \beta ^a e_a\,\right)\,.\label{partlight19}
\end{equation}
If we identify the transversal velocity $\hat{b}^a$ in (\ref{partlight18}) (defined by (\ref{partlight06}) with $v^\alpha = c\,b^\alpha$) with $\beta ^a$ in (\ref{partlight19}) as
\begin{equation}
\hat{b}^a = \beta ^a\,,\label{partlight20}
\end{equation}
in analogy to what we did in (\ref{glob05}), and using (\ref{gamma}), Eq.(\ref{partlight18}) becomes
\begin{equation}
w = c\gamma\,\left(\,\hat{e}_0 + \beta ^a \hat{e}_a\,\right)\,,\label{partlight21}
\end{equation}
having the same components as the r.h.s. of (\ref{partlight19}) but evaluated in a transformed frame, so that $w$ can be regarded as an active transformation of $u$. Referring both, $u$ and $w$, to the rest frame of $u$, see (\ref{partlight19}), such transformation takes the form
\begin{equation}
u = c\,\hat{e}_0\longrightarrow  w = c\gamma\,\left(\,\hat{e}_0 + \beta ^a \hat{e}_a\,\right)\,,\label{activeLor}
\end{equation}
showing the inclination of the trajectory due to the relative velocity (\ref{partlight20}). As in the Galilei case considered above, (\ref{activeLor}) is a particular instance of general active Lorentz transformations of fourvelocities having the same form as (\ref{activeGalbis}).

The factor $\gamma$ present in (\ref{partlight21}) determines that only values $\beta ^2 < 1$ are admissible since otherwise $w$ becomes singular, so that the worldlines of all particle trajectories with timelike vectors $w$ (or $u$) occur inside the light cone (for which $\hat{b}^a =\hat{n}^a$ with $\hat{n}_a \hat{n}^a =1$), the latter constituting the inaccessible boundary of all particle trajectories passing through an event occurring at the cone vertex.

Finally, let us observe that, contrary to what happens in Galilean spacetime (as mentioned at the end of Section IV), in the relativistic case the passive Lorentz transformation of the position (\ref{5dpoinc73}), that is, $p = \hat{\mathfrak{o}} +\hat{\xi}^\alpha\,\hat{e}_\alpha = \mathfrak{o} +\xi ^\alpha\,e_\alpha\,$, implies not only $\hat{\xi}^0 \hat{e}_0$ to become oblique with respect to $\xi ^0 e_0$, but also $\hat{\xi}^a \hat{e}_a$ with respect to $\xi ^a e_a$, as a consequence of (\ref{hate0}) and (\ref{hatea}). The boost-dependent inclination induced by the Lorentz transformations deviates the simultaneity hypersurfaces expanded by (\ref{hatea}) from {\it horizontality}, thus being responsible for simultaneity to be relative rather than absolute.

\section{\bf Comments on dynamics}

Two aspects of geometry are radically new in Einstein's relativistic approaches, accompanying his revision of simultaneity. These are, on the one hand, the inclusion of time in the scheme as a fourth dimension so that only spacetime as a whole makes geometric sense, as assumed by us from the beginning, and on the other hand, the dynamical nature of physical geometry itself.

Indeed, all our previous considerations on the structure of spacetime remain incomplete until connections are determined. In order to fix them, field equations analogous to those of ordinary gauge theories of interactions other than gravity are required \cite{Trautman:1970cy}-\cite{Goeckeler}. In the context of Poincar\'e gauge theories for instance, the field equations can be derived from a principle of extremal action $S=\int L$, being $L=L\left(\,\vartheta ^\alpha \,,R_\alpha{}^\beta \,,T^\alpha \,,\textrm{matter fields}\,\right)$ a Langrange density 4-form depending on tetrads (\ref{partlight01}), curvature (\ref{curvat02}), torsion (\ref{tors02}) and matter fields. The spacetime fields with matter and energy as their sources describe gravity. A detailed discussion of the derivation of Poincar\'e gauge field equations can be found in \cite{Tresguerres:2007ih}.

According to Einstein, the motion equations of a test point particle moving in the dynamically determined spacetime obey the generalized principle of inertia consisting in the vanishing of acceleration (\ref{eqsmot}), that is
\begin{equation}
{\cal \L\/}_u u^\alpha =0\,,\label{noaccel}
\end{equation}
defining autoparallel trajectories. More explicitly, (\ref{noaccel}) is the covariant acceleration (\ref{acceleration}) put equal to zero. According to our geometric interpretation, observable motion consists in the changes of the components $\xi ^\alpha$ of position (\ref{5dpoinc73}) when the latter evolves along a worldline under the action of the operator $\nabla _u$ with vanishing (\ref{eqsmot}), and thus satisfying (\ref{noaccel}). The connections $\Gamma _{\bot\nu}{}^\mu\,$ and ${\buildrel (T)\over{\Gamma _{\bot}^\mu}}$, emerging in (\ref{eqsmot}) due to the evolution of the basis vectors $e_\alpha$ and of the origin $\mathfrak{o}$ respectively, exert influence on the acceleration ${\it{l}}_u {\it{l}}_u\xi ^\alpha$ of the position vector as force-like contributions. See (\ref{acceleration}). In the absence of such connections, (\ref{noaccel}) reduces to the inertial law ${\it{l}}_u {\it{l}}_u\xi ^\alpha =0$, implying rectilinear motion.

In order to illustrate the description of motion in terms of the bundle variables of our approach, let us rework the example, already presented in \cite{Tresguerres:2012nu}, of a test particle moving in a Schwarzschild spacetime. The Schwarzschild solution of General Relativity can be expressed as (\ref{lineelement}), using the Minkowski metric (\ref{Mink}), and tetrads in Cartesian coordinates 
\begin{eqnarray}
\vartheta ^0 &=& \Phi\,d\xi ^0\,,\label{Schwarz04}\\
\vartheta ^a &=& d\xi ^a +\Bigl( {1\over \Phi} -1\Bigr) {{\xi ^a \xi _b}\over{r^2}}\,d\xi ^b\,,\label{Schwarz05}
\end{eqnarray}
with $\xi ^0 = c t$, and being the Schwarzschild function 
\begin{equation}
\Phi :=\sqrt{1-{{2GM}\over{c^2 r}}}\,,\label{Schwarz02}
\end{equation}
where $r =\sqrt{\xi _a \xi ^a} =\sqrt{\delta _{ab}\,\xi ^a \xi ^b}\,$. The tetrads can be rewritten in the form (\ref{partlight01}) with the help of the Christoffel connections
\begin{eqnarray}
\Gamma ^{\{\}}_{0a} &=& {{GM}\over{c^2}}\,{{\xi _a}\over{r^3}}\,d\xi ^0\,,\label{Schwarz08}\\
\Gamma ^{\{\}}_{ab} &=&  (\Phi -1 )\,{{2}\over{r^2}}\,\xi _{ [ a} d\xi _{b ]}\,,\label{Schwarz09}
\end{eqnarray}
and of the linear translational connections
\begin{eqnarray}
{\buildrel (T)\over{\Gamma\,^0}} &=& -{{GM}\over{c^2 r}}\,{{(\Phi +3)}\over{(\Phi +1 )}}\,d\xi ^0\,,\label{Schwarz12}\\
{\buildrel (T)\over{\Gamma\,^a}} &=& (1-\Phi )\Bigr[\,d\xi ^a +\Bigl( {1\over \Phi} -1\Bigr) {{\xi ^a \xi _b}\over{r^2}}\,d\xi ^b\,\Bigr]\nonumber\\
&&-{{GM}\over{c^2}}\,{{\xi ^a \xi ^0}\over{r^3}}\,d\xi ^0\,,\label{Schwarz13}
\end{eqnarray}
(all of them vanishing for zero mass $M$). Contraction of (\ref{Schwarz04}) and (\ref{Schwarz05}) with the {\it time vector} $u$ yields the fourvelocity components (\ref{partlight03}), that is
\begin{eqnarray}
u^0 &=& \Phi\,\dot{\xi}^0\,,\label{Schwarz15}\\
u^a &=& \dot{\xi}^a +\Bigl( {1\over \Phi} -1\Bigr) {{\xi ^a \xi _b}\over{r^2}}\,\dot{\xi}^b\,,\label{Schwarz15bis}
\end{eqnarray}
where the simplified notation $\dot{\xi}^\alpha := {\it{l}}_u \xi ^\alpha = u\rfloor d\xi ^\alpha$ is used. The motion equations (\ref{noaccel}), when applied to (\ref{Schwarz15}) and (\ref{Schwarz15bis}) with the connection components (\ref{Schwarz08}) and (\ref{Schwarz09}), read
\begin{eqnarray}
0= {\cal \L\/}_u\,u^0 &=& {1\over \Phi}\,{\it{l}}_u ( \Phi ^2\,\dot{\xi}^0\,)\,,\label{Schwarz16}\\
0= {\cal \L\/}_u\,u^a &=& \ddot{\xi}^a + {{\xi ^a}\over r}\Bigl\{\,\bigl( {1\over \Phi} -1\bigr)\ddot{r}\nonumber\\
&&\hskip1.7cm + {{(1-\Phi)}\over{r}}\left(\,\dot{\xi}_b \dot{\xi}^b - \dot{r}^2\right)\nonumber\\
&&\hskip1.7cm +{{\partial _r\Phi}\over{\Phi ^2}}\left[\,( \Phi ^2\,\dot{\xi}^0\,)^2 - \dot{r}^2\,\right]\,\Bigr\}\,.\nonumber\\
\label{Schwarz17}
\end{eqnarray}
Eq.(\ref{Schwarz16}) implies
\begin{equation}
\Phi ^2\,\dot{\xi}^0 = K =const\,,\label{Schwarz18}
\end{equation}
while (\ref{Schwarz17}) can be simplified taking into account the relation $u_\alpha u^\alpha =-c^2$ and the fact that the angular momentum per unit mass, $J_a:=\epsilon _{abc}{\xi}^b\dot{\xi}^c$, is a conserved quantity, so that also its square
\begin{equation}
J^2 = r^2 (\,\dot{\xi}_b \dot{\xi}^b -\dot{r}^2 )\label{Schwarz22}
\end{equation}
is a constant. Eq.(\ref{Schwarz17}) reduces to
\begin{equation}
\ddot{\xi}^a = -\,{{GM \xi ^a}\over{r^3}}\left( 1 + {{3 J^2}\over{c^2 r^2}}\right)\,,\label{Schwarz26}
\end{equation}
constituting the general-relativistic Schwarzschild modification of the Newtonian law of gravitation for the position components $\xi ^0 (=: ct)$ and $\xi ^a$ of (\ref{5dpoinc73}).

\section{\bf Conclusions}

We have established a common foundation of different spacetime geometries, making apparent the constructive assumptions leading to each of them. Our starting point was a composite bundle treatment of affine geometry. By imposing a preferred congruence of worldlines to exist, having a time direction ${\buildrel abs\over{u}}$ with its time rate of change ${\buildrel abs\over{u^0}}$ satisfying suitable conditions, we derived the (non necessarily flat) Galilei spacetime of Newtonian mechanics. Alternatively, Einstein's synchronization postulate, together with the time rate condition $\hat{u}^0 =c$ for observers at rest, defines a different preferred family of light worldlines with tangent vector $w_L$, allowing one to construct the Minkowski metric, invariant under local Lorentz transformations. In each case, the condition imposed on $u^0$ determines a different relationship between $\xi ^0$ and $\tau$. So to say, one makes a choice about the meaning of what clock time $\xi ^0$ actually measures.

The Lorentz group, found to be an isometry group of the Minkowski metric, as a subgroup of the general linear group $GL(4\,,\mathbb{R})$, can be accommodated in the original affine scheme in two different ways, both of them compatible with our results. So, either one can adopt the Poincar\'e group as the structure group $G$ of a different fiber bundle, with $H=$Lorentz, giving rise to the geometry of Poincar\'e gauge theories \cite{Hehl:1974cn}-\cite{Obukhov:2006ge} (see Appendix D), or one can rearrange the initial affine bundle with $G$ as the affine group and $H=$Lorentz instead of $H=GL(4\,,\mathbb{R})$, yielding a nonlinear realization of metric-affine geometry with general affine connections including torsion and nonmetricity, but with explicit Lorentz symmetry \cite{Tresguerres:2000qn}.

The pseudo-Riemannian spacetime $(M\,,g)$ of ordinary General Relativity is obtained by considering vanishing torsion and nonmetricity and by pulling back all geometric quantities to the bundle base space $M$, where the line element becomes expressible as $ds^2 = o_{\alpha\beta}\,\vartheta ^\alpha\otimes\vartheta ^\beta = o_{\alpha\beta}\,e_i{}^\alpha dx^i e_j{}^\beta dx^j =: g_{ij}\,dx^i dx^j$. Finally, a geometric structure with vanishing curvature, torsion and nonmetricity, in which connections can be chosen to vanish, reduces to the Minkowskian spacetime of Special Relativity.


\appendix
\section{Affine group}

The affine composite fiber bundle is built starting from the commutation relations of the generators $\lambda ^\alpha{}_\beta$ and $P_\mu$ of the affine group, that is
\begin{eqnarray}
\left[\lambda ^{\alpha}{}_{\beta},  \lambda ^{\mu}{}_{\nu}\right] &=& i\left( \delta^{\alpha}_{\nu}\, \lambda ^{\mu}{}_{\beta} -\delta^{\mu}_{\beta}\,\lambda ^{\alpha}{}_{\nu}\right)\,,\label{affcomrel01}\\
\left[\lambda ^{\alpha}{}_{\beta}\,, P_{\mu}\right] &=& i\,\delta^{\alpha}_{\mu}\, P_{\beta}\,, \label{affcomrel02}\\
\left[P_{\alpha}\,, P_{\beta}\right] &=& 0\,.\label{affcomrel03}
\end{eqnarray}
The group generators admit the 5x5 matrix representation
\begin{eqnarray}
(\lambda ^\alpha{}_\beta )_A{}^B &=&-i\,\delta _A^\alpha\,\delta _{\beta}^B\,,\label{5daffine01}\\
(P_\mu )_A{}^B &=& -i\,l^{-1}\,\delta _A^5\,\delta _\mu ^B\,,\label{5daffine02}
\end{eqnarray}
with $\alpha\,,\beta$ running from $0$ to $3$, and $A\,,B =0,1,2,3,5\,$, and with $l$ as a dimensional constant. The 5x5 representation of $H=GL(4\,,\mathbb{R})$ group elements reads
\begin{equation}
a_A{}^B := ( e^{i\,\zeta _\alpha{}^\beta\,\lambda ^\alpha{}_\beta})_A{}^B = \delta _A^5\delta _5^B + \Lambda _\alpha{}^\beta\,\delta _A^\alpha \delta _\beta ^B\,,\label{5daffmat}
\end{equation}
where
\begin{equation}
\Lambda _\alpha{}^\beta := ( e^{\zeta})_\alpha{}^\beta = \delta _\alpha ^\beta  +\zeta _\alpha{}^\beta +{1\over{2!}}\,\zeta _\alpha{}^\mu \zeta _\mu{}^\beta +{1\over{3!}}\,\zeta _\alpha{}^\mu \zeta _\mu{}^\nu \zeta _\nu{}^\beta +...\,,\label{zeta05}
\end{equation}
while the representation of the translational elements of $G/H$ is given by
\begin{equation}
b_A{}^B := ( e^{-i\,\xi ^\mu P_\mu})_A{}^B =\delta _A^B -l^{-1}\,\xi ^\mu\,\delta _A^5\,\delta _\mu ^B\,.\label{5dtransmat}
\end{equation}
Affine group elements constructed from (\ref{5daffmat}) and (\ref{5dtransmat}) as the product
\begin{eqnarray}
\tilde{g}_A{}^B &=& b_A{}^C\,a_C{}^B\nonumber\\
&=&\delta _A^5\delta _5^B + \Lambda _\alpha{}^\beta\,\delta _A^\alpha \delta _\beta ^B -l^{-1}\,\xi ^\mu \Lambda _\mu{}^\beta \delta _A^5 \delta _\beta ^B \,,\label{affinegroupelem}
\end{eqnarray}
constitute a decomposition characteristic for composite fiber bundles. See Ref.\cite{Tresguerres:2012nu}.

\section{Curvature and torsion}

By acting once more with the operator $\nabla$ on (\ref{5dpoinc78}) and (\ref{5dpoinc80}) respectively, we get
\begin{eqnarray}
\nabla\nabla e_\alpha &=& R_\alpha{}^\beta e_\beta\,,\label{curvat01}\\
\nabla\nabla p &=& T^\alpha e_\alpha\,,\label{tors01}
\end{eqnarray}
where
\begin{equation}
R_\alpha{}^\beta := d \Gamma _\alpha{}^\beta + \Gamma _\gamma{}^\beta\wedge\Gamma _\alpha{}^\gamma\label{curvat02}
\end{equation}
is the spacetime curvature, and
\begin{equation}
T^\alpha := D\vartheta ^\alpha = d\vartheta ^\alpha + \Gamma _\beta{}^\alpha\wedge\vartheta ^\beta\label{tors02}
\end{equation}
stands for torsion.

\section{\bf Galilei transformations}

Let us consider the finite version of the affine transformations (\ref{5dpoinc74})--(\ref{5dpoinc76}), that is
\begin{eqnarray}
\hat{e}_\alpha &=& \Lambda _\alpha{}^\beta e_\beta\,,\label{posit01}\\
\hat{\mathfrak{o}} &=& \mathfrak{o} + a^\mu e_\mu\,,\label{posit02}\\
\hat{\xi}^\alpha &=& (\Lambda ^{-1})_\gamma{}^\alpha\,(\,\xi ^\gamma - a^\gamma )\,,\label{posit03}
\end{eqnarray}
leaving invariant the position (\ref{5dpoinc73}). Galilei transformations are a particular case of them. They can be decomposed into rotations and Galilei boosts as
\begin{equation}
\Lambda _\alpha{}^\beta = \mathfrak{R}_\alpha{}^\gamma\,\mathfrak{B}_\gamma{}^\beta\,,\label{Lormatrix}
\end{equation}
where the matrix elements of the rotational matrix read
\begin{equation}
\mathfrak{R}_0{}^0 = 1\,,\quad\mathfrak{R}_0{}^b = 0\,,\quad\mathfrak{R}_a{}^0 = 0\,,\label{rotmatrix01}
\end{equation}
\begin{equation}
\mathfrak{R}_a{}^b = \cos (|\theta |)\,\delta _a^b + \bigl[\,1 -\cos (|\theta |)\,\bigr]\,{{\theta _a\theta ^b}\over{|\theta |^2}} + \sin (|\theta |)\,{{\theta ^c}\over{|\theta |}}\,\epsilon _{ca}{}^b\,,\label{rotmatrix02}
\end{equation}
with $|\theta | :=\sqrt{\theta _m\theta ^m}\,$, and those of the Galilei boost matrix are
\begin{eqnarray}
&&\mathfrak{B}_0{}^0 = 1\,,\quad\mathfrak{B}_0{}^b = \beta ^b\,,\label{boostmat01}\\
&&\mathfrak{B}_a{}^0 = 0\,,\quad\mathfrak{B}_a{}^b = \delta _a^b\,.\label{boostmat02}
\end{eqnarray}
From (\ref{posit01}) with (\ref{Lormatrix})--(\ref{boostmat02}), we find
\begin{eqnarray}
\hat{e}_0 &=& e_0 +\beta ^a e_a\,,\label{galframe01}\\
\hat{e}_a &=& \mathfrak{R}_a{}^b e_b\,.\label{galframe02}
\end{eqnarray}
Moreover, in terms of the inverse matrices such that
\begin{equation}
(\mathfrak{R}^{-1})_0{}^0 = 1\,,\quad(\mathfrak{R}^{-1})_0{}^b = 0\,,\quad(\mathfrak{R}^{-1})_a{}^0 = 0\,,\label{invrot01}
\end{equation}
and
\begin{eqnarray}
&&(\mathfrak{B}^{-1})_0{}^0 = 1\,,\quad(\mathfrak{B}^{-1})_0{}^b = -\beta ^b\,,\label{boostmat01bis}\\
&&(\mathfrak{B}^{-1})_a{}^0 = 0\,,\quad(\mathfrak{B}^{-1})_a{}^b = \delta _a^b\,,\label{boostmat02bis}
\end{eqnarray}
we get the Galilei transformations (\ref{posit03}) of the coordinate-like position components
\begin{eqnarray}
\hat{\xi}^0 &=& \xi ^0 - a^0\,,\label{galcoord01}\\
\hat{\xi}^a &=& (\mathfrak{R}^{-1})_b{}^a \,[\,\xi ^b - a^b -\beta ^b(\xi ^0 - a^0 ) ]\,.\label{galcoord02}
\end{eqnarray}
Four-velocities transform as covectors
\begin{equation}
\hat{u}^\alpha =(\Lambda ^{-1})_\beta{}^\alpha\,u^\beta\,,\label{galvel01}
\end{equation}
with (\ref{galvel01}) reading explicitly
\begin{eqnarray}
\hat{u}^0 &=& u^0 \,,\label{galvel02}\\
\hat{u}^a &=& (\mathfrak{R}^{-1})_b{}^a \,(\,u^b -\beta ^b u^0 )\,.\label{galvel03}
\end{eqnarray}
Analogously, the transformations of tetrads are
\begin{eqnarray}
\hat{\vartheta}^0 &=& \vartheta ^0 \,,\label{galitetrad02}\\
\hat{\vartheta}^a &=& (\mathfrak{R}^{-1})_b{}^a \,(\,\vartheta ^b -\beta ^b \vartheta ^0 )\,,\label{galitetrad03}
\end{eqnarray}
and those of accelerations, prior to imposing the restriction ${\cal \L\/}_u u^0 =0$ discussed in the main text, read
\begin{eqnarray}
\hat{{\cal \L\/}}_u \hat{u}^0 &=& {\cal \L\/}_u u^0 \,,\label{galaccel02}\\
\hat{{\cal \L\/}}_u \hat{u}^a &=& (\mathfrak{R}^{-1})_b{}^a \,(\,{\cal \L\/}_u u^b -\beta ^b {\cal \L\/}_u u^0 )\,.\label{galaccel03}
\end{eqnarray}
On the other hand, from the composition law of Galilei boosts (\ref{boostmat01}), (\ref{boostmat02}), that is
\begin{equation}
\mathfrak{B}_\alpha{}^\beta (b) = \mathfrak{B}_\alpha{}^\mu (\beta )\mathfrak{B}_\mu{}^\beta (\hat{b})\,,\label{velcomp01}
\end{equation}
one gets the Galilei composition law for velocities
\begin{equation}
b^a = \hat{b}^a +\beta ^a\,,\label{galcomp}
\end{equation}
analogous to (\ref{absoltime06}).

\section{5x5 matrix representation of the Poincar\'e group}

The Lorentz generators $L_{\alpha\beta}$ and the translational generators $P_\alpha\hskip0.1cm (\alpha\,,\beta =\,0\,,...\,,3)\,$ of the Poincar\'e group satisfy the commutation relations
\begin{eqnarray}
\left[L_{\alpha\beta }\,,L_{\mu\nu }\right] &=& -i\,\left( o_{\alpha [\mu } L_{\nu ]\beta} - o_{\beta [\mu }L_{\nu ]\alpha}\right)\,,\label{comrelpoinc01}\\
\left[L_{\alpha\beta }\,, P_\mu\right] &=& i\,o_{\mu [\alpha}P_{\beta ]}\,,\label{comrelpoinc02}\\
\left[P_{\alpha}\, , P_{\beta}\right] &=& 0\,,\label{comrelpoinc03}
\end{eqnarray}
with $o_{\alpha\beta}$ as the the Minkowski metric
\begin{equation}
o_{\alpha\beta}:=diag\,(-+++)\,.\label{metric}
\end{equation}
In analogy to (\ref{5daffine01}), (\ref{5daffine02}), the Poincar\'e group generators have the 5x5 matrix representation
\begin{eqnarray}
( L_{\alpha\beta} )_A{}^B &=&-i\,o_{A[\alpha}\,\delta _{\beta ]}^B\,,\label{5dpoinc01}\\
(P_\mu )_A{}^B &=& -i\,l^{-1}\,\delta _A^5\,\delta _\mu ^B\,.\label{5dpoinc02}
\end{eqnarray}
The object $o_{AB}$ in (\ref{5dpoinc01}) is a symmetric matrix whose components $o_{\alpha\beta}$ are identical with the Minkowski metric (\ref{metric}), and
\begin{equation}
o_{\alpha 5}=0\,,\label{5dpoinc04}
\end{equation}
while the components $o_{55}$, being undetermined, can be fixed at convenience. The 5x5 representation of Poincar\'e group elements are built as in Appendix A. See \cite{Tresguerres:2012nu}.

\end{document}